\documentclass[pra]{revtex4-2}
\usepackage{eurosym}
\usepackage{amssymb}
\usepackage{graphicx}
\usepackage{verbatim}
\usepackage[normalem]{ulem}
\usepackage{color}
\usepackage{amsmath}
\usepackage{amsfonts}

\begin{document}
\title{Stabilization of two-dimensional optical continuous-wave states by a potential trough}

\author{Thawatchai Mayteevarunyoo $^{1}$ and Boris A. Malomed $^{2,3,*}$}
\address{$^1$Department of Electrical and Computer Engineering, Faculty of Engineering, Naresuan University,
Phitsanulok 65000, Thailand\\
$^2$ Department of Physical Electronics, School of Electrical Engineering, Faculty of Engineering, Tel Aviv
University, Tel Aviv 69978, Israel\\
$^3$Instituto de Alta Investigaci\'{o}n, Universidad de Tarapac\'{a}á, Casilla 7D, Arica, Chile\\
$^*$corresponding author}

\begin{abstract}
We consider quasi-one-dimensional (Q1D) continuous waves (CWs) in the
two-dimensional (2D) optical system with the cubic-quintic nonlinearity and
a Q1D potential trough. In the case of a smooth trough profile, we confirm the
known modulational instability (MI) of Q1D CWs with the transverse structure
corresponding to the 1D ground state (GS) in the potential trough, and
demonstrate the MI of CWs with the dipole-mode (DM) transverse structure,
corresponding to the lowest 1D excited state in the potential trough. The
CWs of both GS and DM types remain nearly stable close to edges of their
existence regions. Stable stationary states in the form of periodic chains
of 2D solitons, trapped in the potential trough, are produced in a numerical
form. The dynamics of the soliton chains excited by a localized kick is
studied too. For the potential trough with the singular delta-functional
profile, we find two species of exact analytical solutions for CWs, one of
which is \emph{completely stable}.
\end{abstract}

\maketitle


\section{Introduction and the models}

Spatial-domain propagation of light in bulk waveguides is determined by the
interplay of diffraction and material nonlinearity. This setting gives rise
to a variety of self-supporting modes, including solitons \cite%
{dark,KA,Dauxois}, vortices \cite{OptVort,OptVort2,OptVort3,OptVort4}
(similar to their counterparts in atomic Bose-Einstein condensates (BEC)
\cite{Fetter,Sakaguchi}), skyrmions \cite{Ady,skyrmion}, etc., which are
adequately modeled by the two-dimensional (2D) nonlinear Schr\"{o}dinger
(NLS) equation, or a coupled system of such equations \cite{Sulem,Fibich}.
Spatiotemporal light propagation in planar waveguides is accurately modeled
by 2D NLS equations which are essentially identical to their counterparts in
the spatial domain \cite{old}. While the ubiquitous nonlinearity of optical
materials is represented by the Kerr cubic term, accurate experiments were
also performed in chalcogenide glasses \cite{chalcogenide} and colloidal
suspensions of metallic particles \cite{Cid}, which are adequately
represented by combined nonlinearities -- most typically, cubic-quintic (CQ)
ones, which feature the competition of the cubic self-focusing and quintic
defocusing.

While 1D solitons are typically stable states, the stability is a major
challenge for studies of 2D nonlinear modes in optics, as well as in other
areas of physics, especially in BEC \cite{old,book}. The basic problem is
the \textit{critical collapse} which destabilizes 2D fundamental solitons
under the action of the cubic self-focusing \cite{Sulem,Fibich}, while the
solitons with embedded vorticity are still more vulnerable to the splitting
instability \cite{vortex-unstable}. Two major mechanisms which provide
stabilization of both the fundamental (zero-vorticity) and vortex optical
solitons in 2D rely upon the use of the CQ nonlinearity and/or effective
potentials, which represent the transverse modulation of the refractive
index in the waveguide. In particular, the CQ nonlinearity readily
stabilizes a part of the vortex-soliton states, with different values of the
topological charge (vorticity), against the splitting \cite{Quiroga,Pego}.
On the other hand, spatially periodic (lattice) potentials, both fully
two-dimensional (2D) \cite{BBB,Yang} and quasi-one-dimensional (Q1D) ones,
which depend on a single coordinate \cite{quasi,Mihalache}), also secure the
stabilization of the fundamental and vortex solitons (although the lattice
potentials destroy the underlying spatial isotropy, vortex solitons with a
nonzero topological charge can be defined in this case too \cite{BBB,Yang}).

The most basic 2D\ model combining the CQ nonlinearity and spatial
inhomogeneity includes a Q1D potential trough, which depends on the single
transverse coordinate, $x$, and does not depend on the other coordinate, $y$
\cite{India}. The corresponding scaled NLS equation for the slowly varying
amplitude of the optical wave, $u\left( x,y;z\right) $, is

\begin{equation}
iu_{z}=-\frac{1}{2}\left( u_{xx}+u_{yy}\right) -|u|^{2}u+g|u|^{4}u-W_{0}\exp
\left( -x^{2}\right) u,  \label{NLS}
\end{equation}%
where $z$ is the propagation distance, $g>0$ is the strength of the quintic
self-defocusing nonlinearity (while the strength of the cubic self-focusing
is set to be $1$ by scaling), and $W_{0}>0$ is the depth of the potential
trough with the Gaussian profile, whose width is also fixed to be $1$ by
means of scaling.

Equation (\ref{NLS}) applies as well to the spatiotemporal light propagation
in a planar waveguide with anomalous group-velocity dispersion, In that
case, $x$ remains the transverse coordinate, while $y=t-z/V_{\mathrm{gr}}$
is the temporal variable, with $t$ and $V_{\mathrm{gr}}$ being the scaled
time and group velocity of the carrier wave. In the spatiotemporal setting,
the trough potential represents a stripe of a material with an enhanced
refractive index, embedded into the planar waveguide. In the latter case,
solely the Q1D effective potential is possible, because it cannot be made a
function on time.

Fully localized states, predicted as solutions of Eq. (\ref{NLS}), are
characterized by the value of the 2D integral power in the spatial domain
(or energy in the spatiotemporal one),%
\begin{equation}
P_{\mathrm{2D}}=\int_{-\infty }^{+\infty }dy\int_{-\infty }^{+\infty
}dx\left\vert u\left( x.y\right) \right\vert ^{2},  \label{P}
\end{equation}%
which is a dynamical invariant of Eq. (\ref{NLS}). Other conserved
quantities are the Hamiltonian,
\begin{equation*}
H=\int_{-\infty }^{+\infty }dy\int_{-\infty }^{+\infty }dx\left[ \frac{1}{2}%
\left( \left\vert u_{x}\right\vert ^{2}+\left\vert u_{y}\right\vert
^{2}\right) -\frac{1}{2}|u|^{4}+\frac{g}{3}|u|^{6}-W_{0}\exp \left(
-x^{2}\right) |u|^{2}\right] ,
\end{equation*}%
and the $y$-component of the momentum, $M_{y}=i\int_{-\infty }^{+\infty
}dy\int_{-\infty }^{+\infty }dxu_{y}^{\ast }u$, where $\ast $ stands for the
complex conjugate.

The simplest stationary states admitted by Eq. (\ref{NLS}) are continuous
waves (CWs), which are solutions trapped in the potential trough and thus
localized in the $x$ direction, while their $y$ dependence is represented by
a real wavenumber $q$:%
\begin{equation}
u_{\mathrm{CW}}\left( x,y,z\right) =\exp \left( ikz+iqy\right) U(x).
\label{CW}
\end{equation}%
Here, $k$ is a real propagation constant, and a real modal function $U(x)$
obeys the the transverse equation:%
\begin{equation}
\left( k+\frac{1}{2}q^{2}\right) U=\frac{1}{2}\frac{d^{2}U}{dx^{2}}%
+U^{3}-gU^{5}+W_{0}\exp \left( -x^{2}\right) U.  \label{k}
\end{equation}%
Using the obvious Galilean invariance of Eq. (\ref{NLS}) in the $y$
direction, it is sufficient to consider the CW solutions with $q=0$, which
is done below, the respective CW solution being independent of $y$. Then,
Eq. (\ref{k}) produces bound states (solutions localized at $|x|\rightarrow
\infty $) for $k>0$.

Equation (\ref{k}) is the 1D stationary NLS equation with the Gaussian
trapping potential, Therefore, it is natural to classify its bound-state
solutions as a nonlinear extension of eigenstates which are produced by the
linear Schr\"{o}dinger equation with the trapping potential, in the
small-amplitude limit. Accordingly, the bound states are classified as the
spatially even ($U(-x)=U(x)$) ground state (GS), spatially odd ($U(-x)=-U(x)$%
) first excited state (alias the dipole mode, DM), spatially even second
excited state (alias the quadrupole mode), etc.

The basic problem for CW solutions is their modulational instability (MI,
alias the Benjamin-Feir instability \cite{BF}) against small perturbations
which impose periodic modulation of the CW state along the free coordinate $%
y $ \cite{MI0,MI1,MI2,MI3,MI4,MI5,MI6}. In particular, MI plays the crucial
role in the generation of rogue waves \cite{RW1,RW2,RW3,RW4,RW5}. For the CW
solutions of Eq. (\ref{NLS}) with the GS transverse structure, this problem
was originally addressed in Ref. \cite{India}. In the presents work, we aim
to further elaborate this topic. First, in Section 2 we recapitulate the MI
analysis performed in the framework of Eq. (\ref{NLS}), including new
findings for the CWs with the DM transverse structure. While all the CWs are
subject to MI, the instability is very weak for the CWs of both the GS and
DM types, with values of the propagation constant close to existence
boundaries of these solution families, the respective interval of the
\textquotedblleft practical stability" being wider for the DMs than for GSs.

It is natural to expect that MI may split the CW into a periodic chain of 2D
solitons along the $y$ coordinate, trapped in the potential trough. In
Section 3, we produce a numerical solution for the stationary soliton chain
and demonstrate its stability. We also report results produced by the
application of a kick to a particular soliton in the chain. The kick, that
may be directed along $x$ or $y$, eventually generates a complex dynamical
picture.

The most essential novel results are reported in Section 4 for the 2D model
with the CQ nonlinearity and the potential trough, $W(x)$, represented by
the delta-functional profile:%
\begin{equation}
iu_{z}=-\frac{1}{2}\left( u_{xx}+u_{yy}\right) -|u|^{2}u+g|u|^{4}u-\sqrt{\pi
}W_{0}\delta (x)u.  \label{delta}
\end{equation}%
Factor $\sqrt{\pi }$ is introduced in Eq. (\ref{delta}) to keep the same
transverse area of the trough potential, $\int_{-\infty }^{+\infty }W(x)dx=%
\sqrt{\pi }W_{0}$, as in Eq. (\ref{NLS}). We obtain two families of exact
analytical solutions of Eq. (\ref{delta}), in parameter regions
\begin{equation}
W_{0}<W_{\mathrm{cr}}~~~(\mathbf{Case\ 1})~~~\mathrm{and~~~}W_{0}>W_{\mathrm{%
cr}}~~~(\mathbf{Case\ 2}),  \label{12}
\end{equation}%
where the critical potential strength is
\begin{equation}
W_{\mathrm{cr}}=\sqrt{3/\left( 8\pi g\right) }.  \label{Wcrit}
\end{equation}%
The numerical analysis demonstrates that, while the exact solutions are
subject to MI in \textbf{Case 1}, they are \emph{completely stable} in
\textbf{Case 2}. The possibility to produce stable CW states in the form of
Q1D bright solitons, supported by the trough potential, is a remarkable
finding (in particular, because the stable solutions are obtained here in
the analytical form), which was not reported for previously studied models.
In this connection, it is relevant to mention that it was demonstrated in
Ref. \cite{Ma} that a Q1D (stripe-shaped) \textit{dark soliton} may be
stabilized against MI, in the framework of the two-dimensional NLS equation
with the cubic self-defocusing term (the quintic one is not needed in this
case), by means of a sufficiently strong Q1D \textit{potential ridge}
(rather than the trough), which corresponds to $W_{0}<0$ in Eq. (\ref{NLS}).

The paper is concluded by Section 5, which summarizes the obtained results
and puts forward possibilities for additional studies on the present topic.

\section{MI (modulational instability) of the Q1D (quasi-one-dimensional) CW
states}

\subsection{Linearized equations for modulational perturbations}

To analyze the modulational (in)stability of the CW states, a perturbed
solution is looked for as%
\begin{equation}
u\left( z,x,y\right) =\exp \left( ikz\right) \left\{ U(x)+\varepsilon \left[
a(x)\exp \left( \gamma z+ipy\right) +b^{\ast }(x)\exp \left( \gamma ^{\ast
}z-ipy\right) \right] \right\} ,  \label{pert}
\end{equation}%
where\ $U(x)$ is a real solution of Eq. (\ref{k}) with propagation constant $%
k$ (recall we set $q=0$ in Eqs. (\ref{CW}) and (\ref{k})), $a(x)$ and $b(x)$
are components of the eigenmode of small perturbations, with a real
infinitesimal amplitude $\varepsilon $ and wavenumber $p$, and $\gamma (p)$
is the corresponding eigenvalue, the stability condition being $\mathrm{Re}%
\left( \gamma (p)\right) =0$ for all real $p$. The substitution of ansatz (%
\ref{pert}) in Eq. (\ref{NLS}) and linearization with respect to $%
\varepsilon $ leads to the system of linear equations for the perturbation
eigenmode (the Bogoliubov-de Gennes (BdG) equations, in terms of the BEC
theory \cite{Pit}):
\begin{gather}
\left( -k-\frac{p^{2}}{2}+i\gamma \right) a=  \notag \\
-\frac{1}{2}\frac{d^{2}a}{dx^{2}}-W_{0}\exp \left( -x^{2}\right)
a-2U^{2}(x)a-U^{2}(x)b+3gU^{4}(x)a+2gU^{4}(x)b,  \label{a} \\
\left( -k-\frac{p^{2}}{2}-i\gamma \right) b=  \notag \\
-\frac{1}{2}\frac{d^{2}b}{dx^{2}}-W_{0}\exp \left( -x^{2}\right)
b-2U^{2}(x)b-U^{2}(x)a+3gU^{4}(x)b+2gU^{4}(x)a.  \label{b}
\end{gather}%
The analysis of the CW states and their stability may be performed treating $%
W_{0}>0$ and $g>0$ as control parameters, while $k>0$ is a free parameter of
the family of the CW states.

\subsection{The MI of the CW states with the ground-state (GS) transverse
structure}

In the uniform space, with $W_{0}=0$ in Eq. (\ref{NLS}), the transverse
profile of the CW state is given by the well-known Q1D soliton solution of
Eq. (\ref{NLS}) \cite{Bulgaria}:%
\begin{equation}
u_{\mathrm{sol}}(x,z)=e^{ikz}\sqrt{\frac{2k}{1+\sqrt{1-4gk/3}\,\cosh
\!\left( 2\sqrt{k}\,x\right) }},  \label{PushPush}
\end{equation}%
in which the propagation constant $k$ takes values in the interval
\begin{equation}
0<k<k_{\max }\equiv \frac{3}{4g}.  \label{kmax}
\end{equation}%
The total 1D power of soliton~(\ref{PushPush}),
\begin{equation}
P(k)=\int_{-\infty }^{+\infty }\left\vert u_{\mathrm{sol}}(x)\right\vert
^{2}dx=\frac{\sqrt{3}}{2\sqrt{g}}\ln \!\left( \frac{\sqrt{3}+2\sqrt{kg}}{%
\sqrt{3}-2\sqrt{kg}}\right) ,  \label{Analytical}
\end{equation}%
diverges in the limit of $k\rightarrow k_{\max }$, when the soliton expands
into a CW state in the $x$ direction.

\begin{figure}[h]
\centering\includegraphics[width=2.5in]{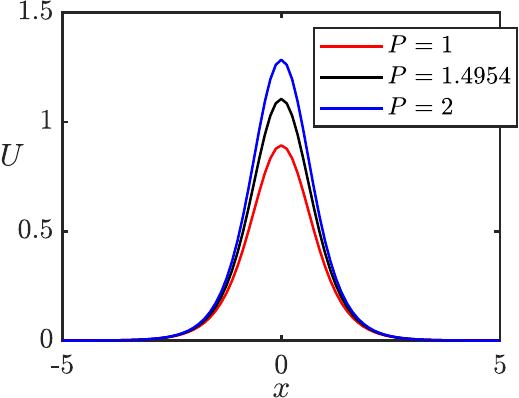}
\caption{Profiles of the GSs produced by the numerical solution of Eq. (%
\protect\ref{k}) with $q=0$, $W_{0}=5.0$, $g=0.5$, and three different
powers, as indicated in the figure. The power $P=1.4954$, with the
respective eigenvalue $k=4.371$ in Eq. (\protect\ref{k}), corresponds to the
largest value of the MI\ gain, $\mathrm{Re}(\protect\gamma )_{\max }=0.8649$%
, obtained from the numerical solution of the BdG system of Eqs. (\protect
\ref{a}) and (\protect\ref{b}).}
\label{Fig1}
\end{figure}

\begin{figure}[h]
\centering\includegraphics[width=3in]{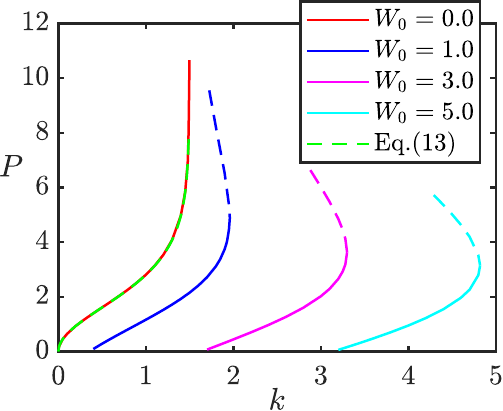}
\caption{Dependences $P(k)$ for the GS families, produced by the numerical
solution of Eq. (\protect\ref{k}) with $q=0$, $g=0.5$, and values of $W_{0}$
indicated in the figure ($W_{0}=0$ pertains to the exact 1D soliton
solution, as given by Eqs. (\protect\ref{PushPush}) and (\protect\ref%
{Analytical})). The 1D power is calculated as $P(k)=\protect\int_{-\infty
}^{+\infty }\left\vert u_{\mathrm{sol}}(x)\right\vert ^{2}dx$, cf. Eq. (%
\protect\ref{Analytical}). Above the turning points, the dashed $P(k)$
curves, if extended towards $P\rightarrow \infty $, approach the value $%
k=3/(4g)$ (see Eq. (\protect\ref{kmax})) from the right.}
\label{Fig2}
\end{figure}

Typical examples of the stationary GSs, produced by the numerical solution
of Eq. (\ref{k}) with $q=0$ and $W_{0}>0$, are plotted in Fig. \ref{Fig1}.
The solutions were produced by means of the modified squared-operator method
(MSOM)\ \cite{MSOM} in the spatial domain $\left( -X/2,+X/2\right) $ of size
$X=40$, discretized on a mesh of $128$ points, with the mesh size $\Delta
x=\allowbreak 0.312\,5$. They may be construed as solitons trapped in the
potential well. For fixed values of parameters $W_{0}$ and $g$, the CW
family is characterized by the dependence of the 1D power $P$ on $k$, cf.
Eq. (\ref{Analytical}) for $W_{0}=0$. In particular, a set of $P(k)$ curves
for $g=0.5$ and different values of $W_{0}$ is plotted in Fig. \ref{Fig2}.
At $P=0$, each curve starts from the eigenvalue which, in terms of quantum
mechanics, is determined by the GS energy, $E_{\mathrm{GS}}\equiv -k(P=0)$,
of the potential well $W(x)=-W_{0}\exp \left( -x^{2}\right) $. With the
increase of of $P$, the curves with $W_{0}>0$ attain their turning points, $%
k=k_{\mathrm{turn}}$, and then move back, in the direction of $k<k_{\mathrm{%
turn}}$, as shown by the dashed lines in Fig. \ref{Fig2}. The extension of
the curves towards $P\rightarrow \infty $ will bring them asymptotically
close to the value $k=3/(4g)$, see Eq. (\ref{kmax}).

\begin{figure}[h]
\centering\includegraphics[width=3in]{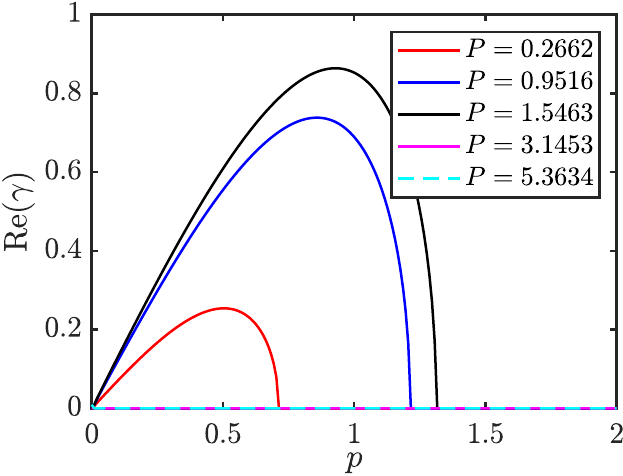}
\caption{The MI gain vs. wavenumber $p$ of the modulational perturbations
(see Eq. (\protect\ref{pert})) for the CW states of the GS type, with $%
W_{0}=5.0$, $g=0.5$, and different values of the 1D power, which are
indicated in the figure. Note that the MI gain is vanishingly small for
large powers, $P=3.1453$ and $\ 5.3634$.}
\label{Fig3}
\end{figure}

In agreement with Ref. \cite{India}, the numerical solution of the BdG
system of Eqs. (\ref{a}) and (\ref{b}) demonstrates that all CW modes of the
GS type are indeed subject to the MI, as the solution always produces
eigenvalues with $\mathrm{Re}\left( \gamma (p)\right) \neq 0$. As an
example, dependences of $\mathrm{Re}\left( \gamma \right) $ on $p$ for $%
W_{0}=5.0$ and $g=0.5$ are plotted in Fig. \ref{Fig3}, for different values
of the 1D power $P$.

\begin{figure}[h]
\centering\includegraphics[width=3in]{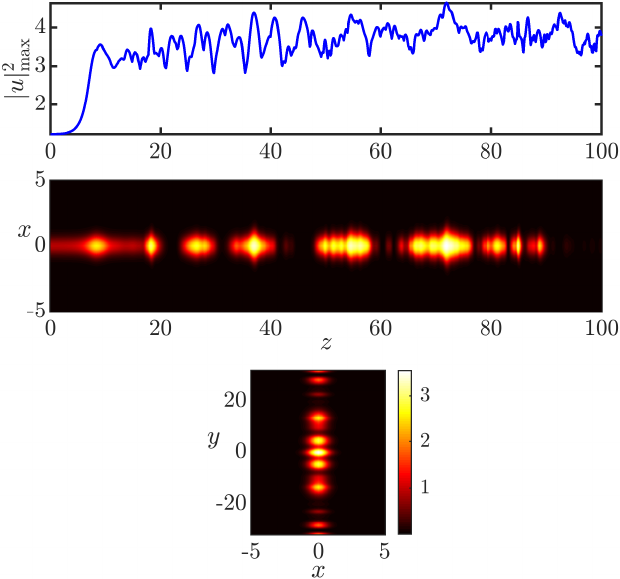}
\caption{The perturbed evolution of the unstable CW state corresponding to
the GS profile with $P=1.4954$ from Fig. \protect\ref{Fig1}, as produced by
simulations of Eq. (\protect\ref{NLS}). Top: the peak intensity $|u|_{\max
}^{2{}}$ vs. $z$, showing rapid growth of the MI, followed by chaotic
oscillations. Middle: the local intensity, $|u|^{2}$, in the $(x,z)$-plane
drawn through $y=0$, showing filamentation of the CW state. Bottom: the
snapshot of the power-density distribution in the $\left( x,y\right) $ plane
at $z=50$, which demonstrates spontaneous splitting of the unstable CW into
a chain of quasi-soliton speckles.}
\label{Fig4}
\end{figure}

The CW\ instability, predicted by the MI analysis, was confirmed by direct
simulations of the perturbed evolution of the CW states, which were carried
out in the framework of Eq. (\ref{NLS}), by means of the usual split-step
Fourier-transform method (here and below, the simulations were performed
with stepsize $\Delta z=0.001$). In particular, results of the simulations
for the CW state with the central GS profile from Fig. \ref{Fig1}, which is
characterized by he largest MI gain, $\mathrm{Re}(\gamma )_{\max }=0.8649$,
for the parameters fixed in Fig. \ref{Fig1}, \textit{viz}., $W_{0}=5.0$ and $%
g=0.5$, are displayed in Fig. \ref{Fig4}. It demonstrates that, in the $%
\left( x,y\right) $ plane, the instability splits the original Q1D CW state
into a chain of localized 2D speckles, which performs irregular
oscillations.

\begin{figure}[h]
\centering\includegraphics[width=3in]{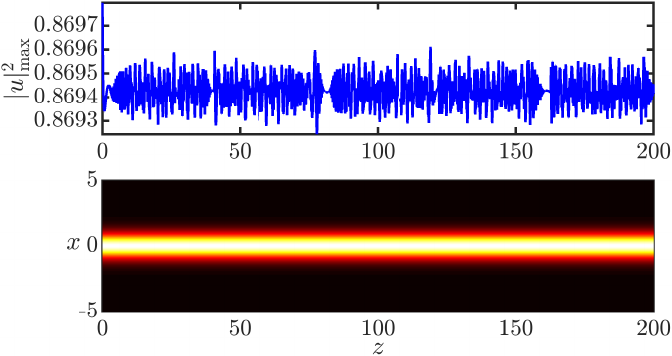}
\caption{The same as in the top and middle panels of Fig. \protect\ref{Fig2}%
, but for a \textquotedblleft practically stable" CW, corresponding to $%
W_{0}=2.5$. $g=1.5,$ and $P=1.3395,~k=1.845$. The actual stability of this
CW is explained by a very small MI gain, as obtained from the numerical
solution of the BdG system of Eqs. (\protect\ref{a}) and (\protect\ref{b}): $%
\mathrm{Re}(\protect\gamma )_{\max }^{\mathrm{peak}}=0.0343$.}
\label{Fig5}
\end{figure}

The CW states corresponding to other values of eigenvalue $k$ in Eq. (\ref{k}%
), which are taken closer to the linear limit, i.e., lower edge of the
respective $P(k)$ curve (see Fig. \ref{Fig2}), demonstrate much weaker MI,
so that, in direct simulations, they may seem as \emph{practically stable}
states. An example is presented in Fig. \ref{Fig5}, for $W_{0}=2.5$, $g=1.5$%
, and $P=1.3395$, $k=1.845$, It is obvious that this CW state, for which the
numerical solution of the BdG equations (\ref{a}) and (\ref{b}) yields a
very small peak value of the MI gain, $\mathrm{Re}(\gamma )_{\max }=0.0343$,
indeed exhibits virtually no instability. On the other hand, CWs with large
values of $P$, such as $P=P=3.1453$ and $\ 5.3634$ in Fig. \ref{Fig3}, may
also demonstrate practical stability, explained by the domination of the
self-defocusing quintic nonlinearity in this case.

\subsection{The MI of the dipole-mode (DM) CWs}

CW states of the DM type were not studied in previous works in the framework
of Eqs. (\ref{NLS}) and (\ref{k}), or in similar models (cf. Ref. \cite%
{India}). Note that, unlike the 1D soliton (\ref{PushPush}), Eq. (\ref{NLS})
in the free space (with $W_{0}=0$) does not have solutions of the DM type.

\begin{figure}[h]
\centering\includegraphics[width=4in]{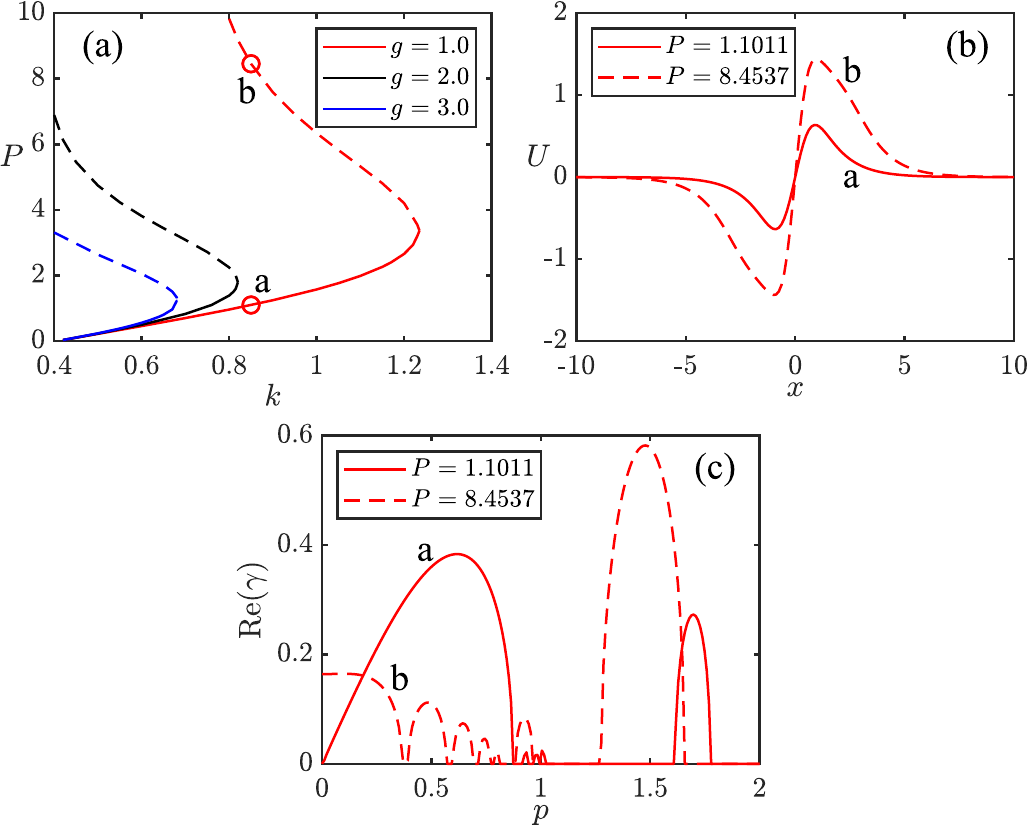}
\caption{(a) $P(k)$ curves for DM families with $W_{0}=5.0$ and three values
of $g$ indicated in the panel, cf. Fig. \protect\ref{Fig2} for the GS modes.
Circles \textrm{a} and \textrm{b} designate DM solutions for $g=1,$ with two
widely diffrent values of the 1D power, $D=1.1011$ and $8.4537$. (b)
Profiles of the solutions designated in (a). (c) The MI gain vs. wavenumber $%
p$ of the modulational perturbations (see Eq. (\protect\ref{pert})) for the
same CW solutions which are designated in panel (a) and presented in (b).}
\label{Fig6}
\end{figure}

Families of numerically found stationary CWs of the DM type are represented
by the respective $P(k)$ curves in Fig. \ref{Fig6}(a) for $W_{0}=5.0$ and
different values of $g$, where $P=\int_{-\infty }^{+\infty }\left\vert
u(x)\right\vert ^{2}dx$ is the same 1D power as defined above. Like their GS
counterparts (cf. Fig. \ref{Fig2}), the curves originate, at $P=0$, from
eigenvalues $k$ corresponding to the first excited state in the
quantum-mechanical potential $W(x)=W_{0}\exp \left( -x^{2}\right) $, reach
turning points at $k=k_{\mathrm{turn}}$, and with further increase of $P$
move back in the direction of $k<k_{\mathrm{turn}}$, as shown by dashed
lines in Fig. \ref{Fig6}(a). Figure \ref{Fig5}(b) demonstrates the DM
profiles for two widely different values of the 1D power, $P=1.1011$ and $%
8.4537$, which corresponds to markers \textrm{a} and \textrm{b} in panel
(a), for $g=1.0$. Lastly, Fig. \ref{Fig5}(c) exhibits spectra of the MI gain
for these CW states.

\begin{figure}[h]
\centering\includegraphics[width=4in]{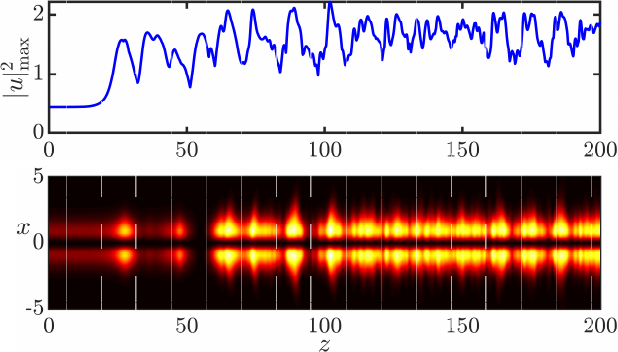}
\caption{The same as in the top and middle panels of Fig. \protect\ref{Fig4}%
, but for the perturbed evolution of the CW of the DM types, in the case of $%
W_{0}=5.0$, $g=1.0$. and $P=1.2$, the corresponding propagation constant
being $k=0.8843.$ For this CW state, the MI gain is $\mathrm{Re}(\protect%
\gamma )_{\max }=0.4012$, attained at the wavenumber of the modulational
perturbation $p=0.8945$, see Eq. (\protect\ref{pert}).}
\label{Fig7}
\end{figure}

\begin{figure}[h]
\centering\includegraphics[width=4in]{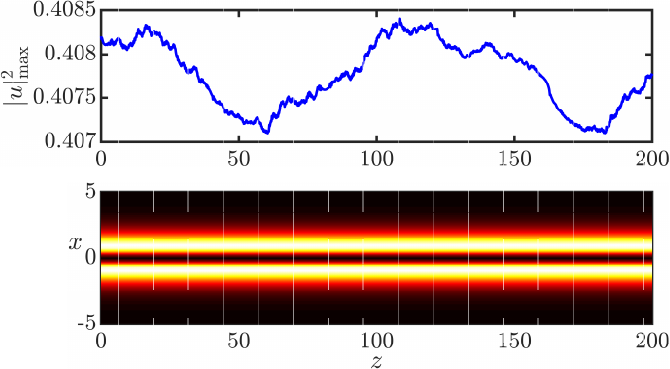}
\caption{The same as in Fig. \protect\ref{Fig7}, but for the perturbed
evolution of the \textquotedblleft practically stable" CW of the DM type, in
the case of $W_{0}=5.0$, $g=3.0$, and $P=1.2$ (the same value of $P$ as in
Fig. \protect\ref{Fig7}), he corresponding propagation constant being $%
k=0.6817.$ For this CW state, the MI gain is $\mathrm{Re}(\protect\gamma %
)_{\max }=0.0188$, attained at the wavenumber of the modulational
perturbation $p=0.1508$, see Eq. (\protect\ref{pert}).}
\label{Fig8}
\end{figure}

In agreement with the predictions of the MI analysis, direct simulations of
the perturbed evolution of the CW states of the DM type demonstrate strong
instability when the respective MI gain,\ $\mathrm{Re}(\gamma )_{\max }$, is
relatively large, and \textquotedblleft practical stability" when the MI
gain is small, see typical examples displayed in Figs. \ref{Fig7} and \ref%
{Fig8}, respectively. The strong MI splits the flat CW into a randomly
oscillating chain of localized speckles, which keep the original dipole
structure. It is relevant to stress that the region of the \textquotedblleft
practical stability" for the CW states of the DM type is essentially broader
than for their GS\ counterparts. Similar to what is concluded above for the
CW states of the GS type, the effective stabilization for the DM CWs may be
provided, as in Fig. \ref{Fig8}, by a sufficiently large value of strength $%
g $ of the defocusing term ($g=3$ in Fig. \ref{Fig8}), or by a large value
of the 1D power, which also makes the defocusing nonlinear term a dominant
one. For instance, in the same case of $W_{0}=5.0$ and $g=1.0$ as in Fig. %
\ref{Fig7}, but with the power larger by a factor $\sim 3$, the CW of the DM
types seems to be fully stable in direct simulations (not shown here in
detail)..

We have also investigated CW states which correspond, in the linear limit,
to the second-order (quadrupole) bound state in the trapping potential $%
W(x)=W_{0}\exp \left( -x^{2}\right) $. They always demonstrate strong MI,
without any interval of the \textquotedblleft practical stability",
therefore they are not considered in detail here.

\section{Stationary chains of 2D solitons}

\subsection{The existence and stability of the chains}

While, as shown in Fig. \ref{Fig4}, the MI splits the unstable CW states
with the GS transverse structure into a chain of localized speckles in the
state of random oscillations, it is natural to expect that the same Eq. (\ref%
{NLS}) may produce a stationary chain of 2D solitons, arranged periodically
along the $y$ axis. Such a chain may be considered as a manifestation of the
MI in terms of stationary states.

Here, we report numerical results for the chains, systematically collected
for parameters $W_{0}=3$ and $g=1$ in Eq. (\ref{NLS}), which makes it
possible to exhibit generic findings. The stationary-chain solutions with
propagation constant $k$ were looked for as%
\begin{equation}
u\left( x,y,z\right) =\exp \left( ikz\right) U\left( x,y\right) ,  \label{2D}
\end{equation}%
with real function $U(x,y)$ satisfying the equation (written with the
above-mentioned values $W_{0}=3$, $g=1$)
\begin{equation}
kU=\frac{1}{2}\left( \frac{\partial ^{2}U}{\partial x^{2}}+\frac{\partial
^{2}U}{\partial y^{2}}\right) +U^{3}-U^{5}+3\exp \left( -x^{2}\right) U,
\label{k2D}
\end{equation}%
cf. Eq. (\ref{k}). The chain solutions were obtained by means of MSOM
applied to Eq. (\ref{k2D}), starting with the $y$-periodic input,
\begin{equation}
U_{0}(x,y)=A_{0}\exp \!\left( -\frac{x^{2}}{4}\right) \cos \!\left( \frac{%
\pi y}{2}\right) .  \label{U0}
\end{equation}%
The numerical solutions were constructed in the computational domain $%
-32\leq x,y\leq +32$ with $256\times 256$ grid points, periodic boundary
conditions in the $\ y$ direction, and zero boundary conditions at $x=\pm 32$%
. The same boundary conditions were adopted in dynamical simulations of Eq. (%
\ref{NLS}).

\begin{figure}[h]
\centering\includegraphics[width=2.5in]{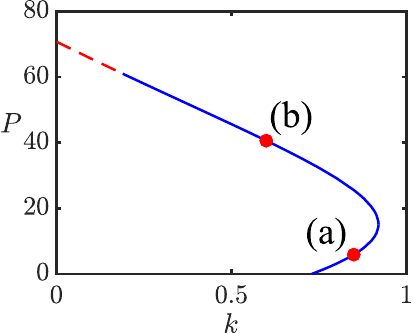}
\caption{Two-dimensional power $P$ of the stationary soliton-chain solutions
vs. the propagation constant $k$, as produced by the numerical solution of
Eq. (\protect\ref{k2D}), with the turning point at $k\approx 0.921$. The
family is stable, except for the red-dashed part at small values of $k$ and
large $P$. Markers~(a) and~(b) indicate the representative solutions, with $%
k_{a}=0.85$ and $k_{b}=0.60$, respectively, which are displayed in Fig.~%
\protect\ref{Fig10}.}
\label{Fig9}
\end{figure}

The soliton-chain family, produced by the numerical solution of Eq. (\ref%
{k2D}), is presented in Fig.~\ref{Fig9} by means of the dependence of the 2D
integral power (\ref{P}) on the propagation constant $k$. The curve features
a turning point at $k\approx 0.921$, which connects two branches with
opposite slopes. The bottom branch originates, at $P=0$, from $k=k_{\mathrm{%
max}}\approx 0.735$, which corresponds to the eigenstate of the respective
2D linear Schr\"{o}dinger equation. Stability of the soliton chains\
belonging to the continuous blue part of the $P(k)$ curve in Fig. \ref{Fig9}
was established by the computation of the respective eigenvalues $\gamma $,
as per the BdG system of Eqs. (\ref{a}) and (\ref{b}), and verified by
direct simulations of the perturbed propagation.

\begin{figure}[h]
\centering\includegraphics[width=4in]{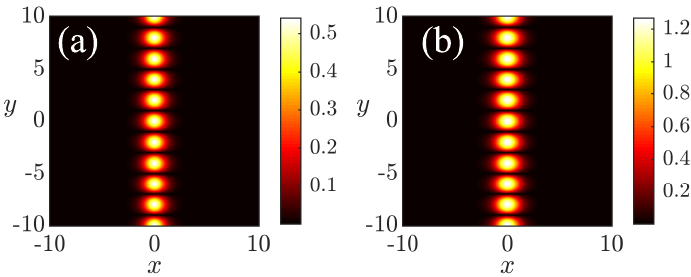}
\caption{Examples of stable soliton-chain solutions, labeled by (a) and (b)
in Fig. \protect\ref{Fig9}. In panels (a) and (b) the solutions correspond
to the propagation constant, total power, and amplitude $k_{\mathrm{a}}=0.85$%
, $P_{\mathrm{a}}=5.8875$, $\left( U_{\max }\right) _{\mathrm{a}}=0.5405$
and $k_{\mathrm{b}}=0.60$, $P_{\mathrm{b}}=40.6236$, $\left( U_{\max
}\right) _{\mathrm{b}}=1.2666$, respectively.}
\label{Fig10}
\end{figure}

Two representative chain solutions, belonging to the bottom and top branches
of the $P(k)$ curve, which are marked (a) and~(b) in Fig.~\ref{Fig9}, are
displayed in Fig. \ref{Fig10}, by means of the respective distributions of $%
\left\vert U\left( x,y\right) \right\vert $, with amplitudes $\left( U_{\max
}\right) _{\mathrm{a}}=0.5405$ and $\left( U_{\max }\right) _{\mathrm{b}%
}=1.2666$ (in agreement with the fact that the power at point (b) in Fig. %
\ref{Fig9} is essentially higher than at point (a)). Adjacent peaks in both
chains keep alternating signs, inherited from the input given by Eq. (\ref%
{U0}).

\subsection{Response of the soliton chains to localized kicks}

To further examine the robustness of the solitons chains and their intrinsic
dynamics, we applied a localized phase kick to the single soliton, whose
center is located at $\left( x,y\right) =\left( 0,0\right) $ in the stable
chain marked (b) in Fig. \ref{Fig9} (the stationary shape of the chain is
plotted in Fig. \ref{Fig10}(b)). The evolution excited by the kick was
observed in direct simulations of Eq. (\ref{NLS}).

First, the kick is applied in the $y$ direction, which is represented by\
the initial condition
\begin{equation}
u(x,y;z=0)=U(x,y)\left[ 1+\left( e^{iQy}-1\right) \xi (x,y)\right] ,
\label{eq:kick}
\end{equation}%
where $U(x,y)$ is the stationary soliton-chain solution,
\begin{equation}
\xi (x,y)=\exp \!\left( -\frac{x^{2}}{\sigma _{x}^{2}}-\frac{y^{2}}{\sigma
_{y}^{2}}\right)  \label{eq:mask}
\end{equation}%
is the 2D Gaussian mask, with widths $\sigma _{x}$ and $\sigma _{y}$ matched
to the transverse dimensions of the individual soliton in the chain, and
real $Q$ is the strength of the kick. Here, we display the results for $%
Q=0.9 $, which represents a generic case.

The results of the simulations, up to $z=120$, are displayed in Fig.~\ref%
{Fig11}. The chain structure remains essentially undisturbed at $z\lesssim
40 $, which is followed by the emission\ of a disturbance wave from the
kicked soliton in the $\pm y$ directions. Due to the periodic boundary
conditions, the disturbances re-enter from the opposite edges of the
computational domain, making the cumulative effect visible at $z\approx 60$.
As a result, an irregular pattern appears at $z\gtrsim 80$. This effect
depends on the size (period) of the domain in the $y$ direction, as
confirmed by simulations performed in the domain with the double size, $%
|y|\leq 64$: the onset of the complex pattern is delayed, and the
disturbance waves re-enter from the opposite edge at $z\approx 120$ (not
shown here in detail).

\begin{figure}[h]
\centering\includegraphics[width=4in]{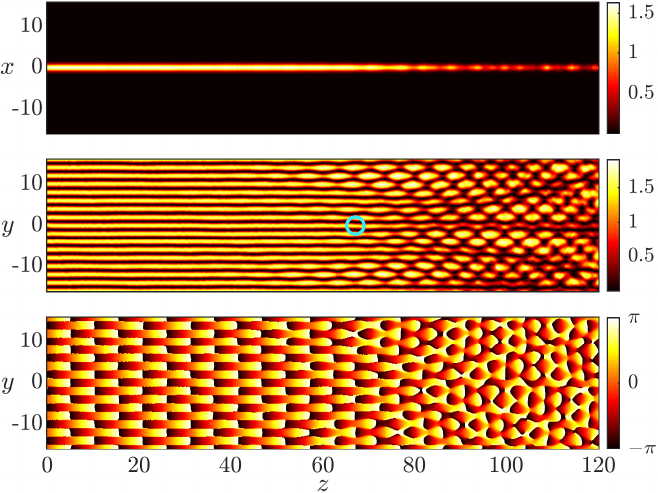}
\caption{The response of the soliton chain, whose stationary form, with $%
k=0.60$ and $P=40.6236$, is plotted in Fig. \protect\ref{Fig10}(b), to the
localized kick applied in the $y$ direction, as per Eq. (\protect\ref%
{eq:kick}) with $Q=0.9$. Top: the evolution of the local intensity $%
\left\vert u(x,z)\right\vert ^{2}$ in the cross-section $y=0$. Middle: the
evolution of $|u\left( y,z\right) |^{2}$ in the cross-section $x=0$; the
cyan circle marks the location of the kicked soliton. Bottom: the evolution
of the phase of $u(y,z)$ in the cross section $x=0$.}
\label{Fig11}
\end{figure}

\begin{figure}[h]
\centering\includegraphics[width=4in]{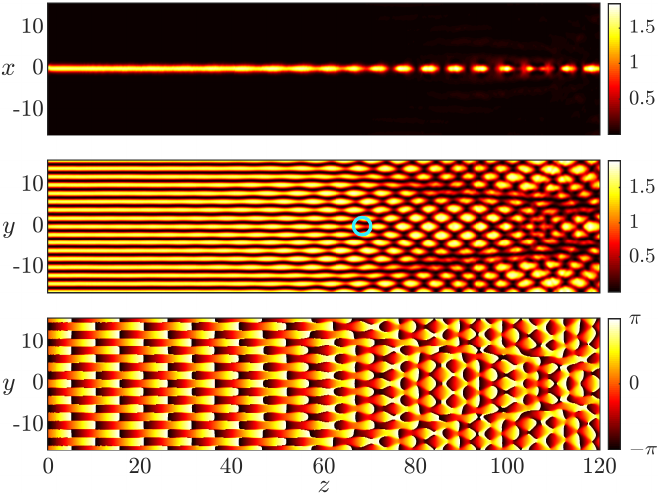}
\caption{The same as in Fig.~\protect\ref{Fig11}, but for the localized kick
applied in the $x$-direction, as per Eq. (\protect\ref{kick-x}) with $Q=0.9$%
. }
\label{Fig12}
\end{figure}

Figure~\ref{Fig12} displays the results produced by the application of the
kick in the $x$ direction, defined by the initial condition%
\begin{equation}
u(x,y;z=0)=U(x,y)\left[ 1+\left( e^{iQx}-1\right) \xi (x,y)\right] ,
\label{kick-x}
\end{equation}%
again with $Q=0.9$. The ensuing picture is qualitatively similar to that in
Fig. \ref{Fig11}, which was initiated by the kick applied in the $y$
direction.

\section{CWs maintained by the delta-functional trapping potential}

\subsection{Analytical results}

The substitution of the usual ansatz for the stationary CW state, $u\left(
x,y,z\right) =\exp \left( ikz\right) U(x)$, in Eq. (\ref{delta}) leads to
the real equation for $U(x)$, cf. Eq. (\ref{k}):%
\begin{equation}
kU=\frac{1}{2}\frac{d^{2}U}{dx^{2}}+U^{3}-gU^{5}+\sqrt{\pi }W_{0}\delta (x)U.
\label{kdelta}
\end{equation}%
The GS solution to Eq. (\ref{kdelta}) for an even function $U(x)$ must be
continuous at $x=0$, with the jump of the first derivative, imposed by the
delta-functional potential:%
\begin{equation}
\frac{dU}{dx}\left( x=+0\right) -\frac{dU}{dx}\left( x=-0\right) =-2\sqrt{%
\pi }W_{0}U(x=0).  \label{jump}
\end{equation}%
A commonly known fact is that the linearized version of Eq. (\ref{kdelta})
supports precisely one bound state, in the form of%
\begin{equation}
U(x)=U_{0}\exp \left( -\sqrt{\pi }W_{0}|x|\right) \mathrm{,~with}~~k=(\pi
/2)W_{0}^{2}~.  \label{linear}
\end{equation}

It is possible to construct two \emph{exact solutions} for the soliton
pinned to the delta-functional potential. First, in \textbf{Case 1}, defined
as per Eqs. (\ref{12}) and (\ref{Wcrit}), i.e., for $W_{0}<\sqrt{3/\left(
8\pi g\right) }$, the exact solution can be constructed in the following
interval of values of the propagation constant,%
\begin{equation}
(\pi /2)W_{0}^{2}\leq k<3/\left( 16g\right) ,  \label{interval1}
\end{equation}%
where the left edge naturally coincides with the eigenvalue (\ref{linear})
of the linearized equation (\ref{kdelta}). If condition (\ref{interval1})
holds, one easily finds a solution, combining the usual exact soliton (\ref%
{PushPush}) in the free space (with $W_{0}=0$) and the jump condition (\ref%
{jump}):%
\begin{equation}
U_{\mathrm{sol}}^{(1)}\left( x,\xi _{1}\right) =\sqrt{\frac{4k}{1+\sqrt{1-%
\frac{16}{3}gk}\cosh \left( 2\sqrt{2k}\left( |x|+\xi _{1}\right) \right) }},
\label{sol1}
\end{equation}%
with the spatial shift $\xi _{1}>0$ determined as a root of equation%
\begin{equation}
\sqrt{2k\left( 1-\frac{16}{3}gk\right) }\sinh \left( 2\sqrt{2k}\xi \right) =%
\sqrt{\pi }W_{0}\left[ 1+\sqrt{1-\frac{16}{3}gk}\cosh \left( 2\sqrt{2k}\xi
\right) \right] .  \label{eq1}
\end{equation}%
An explicit solution of Eq. (\ref{eq1}) is given by a cumbersome expression:%
\begin{equation}
\xi _{1}=\frac{1}{2\sqrt{2k}}\ln \left[ \frac{\sqrt{1+\left( 1-\frac{16}{3}%
gk\right) \left( \frac{2k}{\pi W_{0}^{2}}-1\right) }+1}{\sqrt{1-\frac{16}{3}%
gk}\left( \frac{\sqrt{2k}}{\sqrt{\pi }W_{0}}-1\right) }\right] .  \label{xi1}
\end{equation}

In the opposite \textbf{Case 2}, also defined as per Eqs. (\ref{12}) and (%
\ref{Wcrit}), i.e., as $W_{0}>\sqrt{3/\left( 8\pi g\right) }$, another exact
solution for the soliton pinned to the delta-functional potential can be
found in the following interval of values of the propagation constant:%
\begin{equation}
3/\left( 16g\right) <k\leq (\pi /2)W_{0}^{2},  \label{interval2}
\end{equation}%
cf. interval (\ref{interval1}). In this case, the right edge of the
existence interval coincides with the eigenvalue (\ref{linear}) of the
linearized equation (\ref{kdelta}). The solution existing in interval (\ref%
{interval2}) is based on the following \emph{singular} exact solution of Eq.
(\ref{kdelta}) in the free space (with $W_{0}=0$),%
\begin{equation}
U_{\mathrm{sing}}(x)=\sqrt{\frac{4k}{1+\sqrt{\frac{16}{3}gk-1}\sinh \left( 2%
\sqrt{2k}x\right) }}.  \label{sing}
\end{equation}%
In the free space, the singular solution is irrelevant. However, in the
presence of the delta-functional potential, an exact solution for the pinned
soliton can be obtained from expression (\ref{sing}), replacing $x$ by $%
|x|+\xi _{2}$:%
\begin{equation}
U_{\mathrm{sol}}^{(2)}\left( x,\xi _{2}\right) =\sqrt{\frac{4k}{1+\sqrt{%
\frac{16}{3}gk-1}\sinh \left( 2\sqrt{2k}\left( |x|+\xi _{2}\right) \right) }}
\label{sol2}
\end{equation}%
(cf. expression (\ref{sol1}) for the first soliton solution), with $\xi _{2}$
defined by the equation%
\begin{equation}
\sqrt{2k\left( \frac{16}{3}gk-1\right) }\cosh \left( 2\sqrt{2k}\xi
_{2}\right) =\sqrt{\pi }W_{0}\left[ 1+\sqrt{\frac{16}{3}gk-1}\sinh \left( 2%
\sqrt{2k}\xi _{2}\right) \right] ,  \label{eq2}
\end{equation}%
cf. Eq. (\ref{eq1}) for the first soliton solution. The explicit solution of
Eq. (\ref{eq2}) is%
\begin{equation}
\xi _{2}=\frac{1}{2\sqrt{2k}}\ln \left[ \frac{\sqrt{1+\left( \frac{16}{3}%
gk-1\right) \left( 1-\frac{2k}{\pi W_{0}^{2}}\right) }-1}{\sqrt{\frac{16}{3}%
gk-1}\left( 1-\frac{\sqrt{2k}}{\sqrt{\pi }W_{0}}\right) }\right] ,
\label{xi2}
\end{equation}%
cf. Eq. (\ref{xi1}). Note that, unlike positive $\xi _{1}$ given by Eq. (\ref%
{xi1}) for the first soliton solution, Eq. (\ref{xi2}) may produce $\xi
_{2}<0$.

\subsection{Numerical results}

The accuracy and stability of the above analytical solutions were verified
by comparison with numerical solutions obtained from Eqs. (\ref{delta}) and (%
\ref{kdelta}) with the delta-function approximated by a narrow Gaussian,
\begin{equation}
\delta (x)\rightarrow \tilde{\delta}(x)=\frac{1}{\sqrt{\pi }x_{0}}\exp
\left( -\frac{x^{2}}{x_{0}}\right) ,  \label{approx}
\end{equation}%
with a small regularization scale $x_{0}=(2\Delta x)^{2}$, where $\Delta
x=\allowbreak 0.312\,5$. is the above-mentioned mesh size adopted in
realization of the MSOM algorithm. This value is definitely much smaller
than characteristic length scales, in the $x$ direction, of all solutions
considered here. However, this approximation cannot exactly reproduce the
cusp at the center of exact analytical solutions (\ref{sol1}) and (\ref{sol2}%
) for the pinned solitons, see Fig. \ref{Fig14} below.

\begin{figure}[h]
\centering\includegraphics[width=4in]{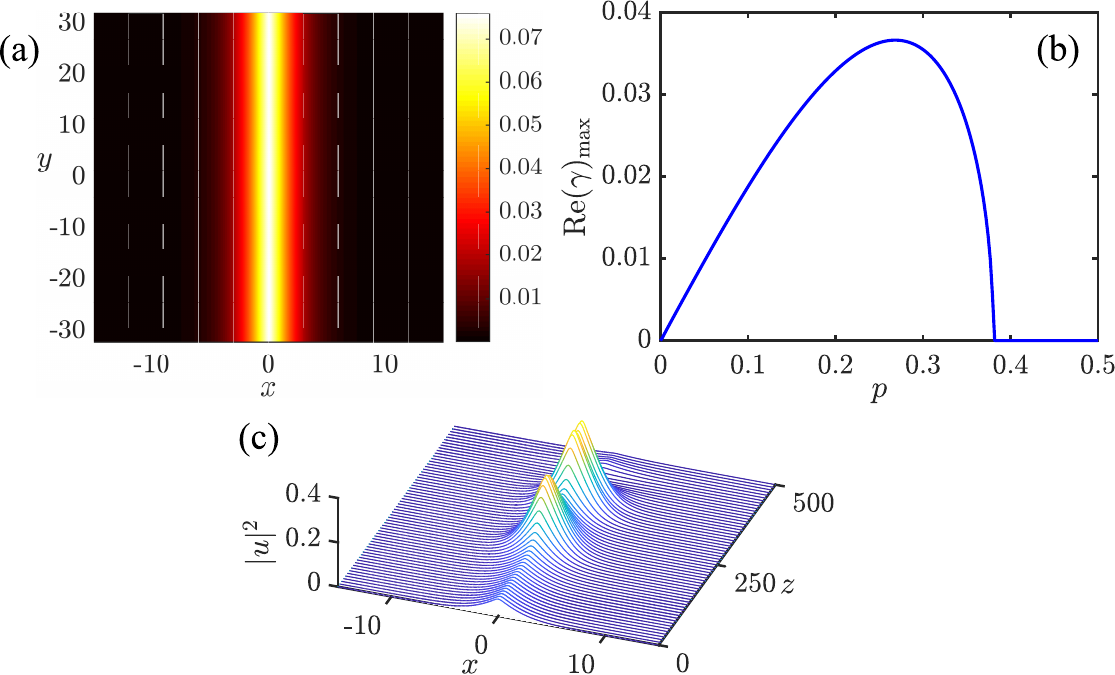}
\caption{An example of the MI destroying the CW based on the stationary
profile (\protect\ref{sol1}) (corresponding to \textbf{Case 1}, see Eq. (%
\protect\ref{12})). The parameters are $W_{0}=0.1$, $g=2$, $k=0.05$. (a)~The
initial intensity profile $|u(x,y;z=0)|^{2}$. (b)~The MI gain, $\mathrm{Re}(%
\protect\gamma )$, versus $p$, showing the instability band with the peak
value $\mathrm{Re}(\protect\gamma )_{\max }\approx 0.037$, attained at $%
p\approx 0.27$. (c)~The evolution of the local intensity $|u(x,y{=}0,z)|^{2}$
in cross-section $y=0$, initiated by the perturbed input (\protect\ref{eps})
with $p=0.2688$ and $\protect\varepsilon =10^{-3}$. The CW breakup commences
at $z\approx 150$.}
\label{Fig13}
\end{figure}

First, both the numerical computation of the stability eigenvalues, as per
the BdG system of Eqs. (\ref{a}) and (\ref{b}), and direct simulations of
the perturbed evolution easily demonstrate that, in terms of the 1D
reduction of Eq. (\ref{delta}) (excluding coordinate $y$), all pinned states
of both types, (\ref{sol1}) and (\ref{sol2}), are completely stable.
However, in the framework of the full 2D equation (\ref{delta}), all the CW
states based on the 1D solitons of type (\ref{sol1}) are subject to MI, see
a typical example in Fig. \ref{Fig13}. In particular, the direct test of the
perturbed propagation, displayed in Fig. \ref{Fig13}(c), was performed by
simulating Eq. (\ref{kdelta}) with the initial condition%
\begin{equation}
u(x,y;z=0)=U(x)\left[ 1+2\varepsilon \cos (py)\right] ,  \label{eps}
\end{equation}%
where $U(x)$ is the stationary profile, $p$ and $\varepsilon $ being the
wavenumber and amplitude of the modulational perturbation (cf. Eq. (\ref%
{pert})).

In view of the full instability of the CWs with the stationary profile (\ref%
{sol1}), we do not consider this case in further detail, and focus on the
CWs with the transverse profile (\ref{sol2}), which turn out to be \emph{%
completely stable}. First, an example of this profile for $W_{0}=1$, $g=1$
and $k=1.3$ is plotted in Fig. \ref{Fig14}.
\begin{figure}[h]
\centering\includegraphics[width=3in]{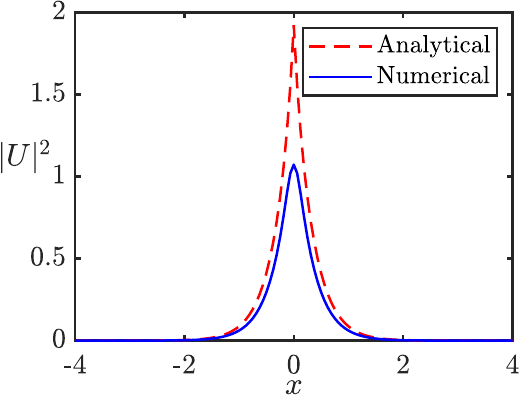}
\caption{An example of the transverse profile of a modulationally stable CW.
The analytical profile is produced by Eqs. (\protect\ref{sol2}) and (\protect
\ref{xi2}), with $W_{0}=g=1$ and $k=1.3$. \ Its numerically found
counterpart cannot exactly reproduce the cusp of the analytical solution,
determined by the jump condition (\protect\ref{jump}).}
\label{Fig14}
\end{figure}
Further, a family of such profiles, with the same parameters $W_{0}=1$, $g=1$
and varying values of $k$, is displayed in Fig. \ref{Fig15}, by means of the
respective dependence of the 1D power $P$ on the propagation constant $k$.
In fact, the power of the analytical solution diverges in the limit of $%
k\rightarrow 3/\left( 16g\right) $, which represents the left edge of the
existence region in Fig. \ref{Fig15}.
\begin{figure}[h]
\centering\includegraphics[width=3in]{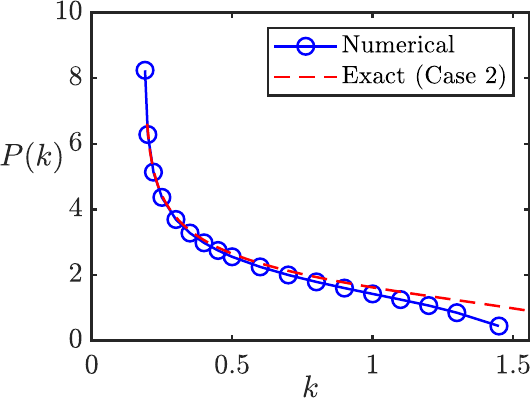}
\caption{The family of the transverse CW profiles, produced by Eq. (\protect
\ref{kdelta}), with $W_{0}=g=1$, which corresponds to \textbf{Case 2}, in
terms of Eq. (\protect\ref{12}). The family is represented by the dependence
of the 1D power $P$ on $k$. The dashed red curve shows the result generated
by the integration of the exact solution (\protect\ref{sol2}), while the
chain of circles shows the values of the power for the numerically found
solutions.}
\label{Fig15}
\end{figure}
Note that the curve $P(k)$ in Fig. \ref{Fig15} obeys the \textit{%
anti-Vakhitov-Kolokolov }(anti-VK) criterion, $dP/dk<0$, which is known as
the necessary stability condition for self-trapped states dominated by
self-repulsive nonlinearity \cite{HS}, which is the case here, because the
pinned solitons of type (\ref{sol2}) do not exist in the absence of the
self-defocusing quintic term in Eq. (\ref{delta}) (the VK criterion proper, $%
dP/dk>0$, is the necessary stability condition for solitons supported by
self-attractive nonlinearity \cite{Vakh,Berge})

\begin{figure}[h]
\centering\includegraphics[width=4in]{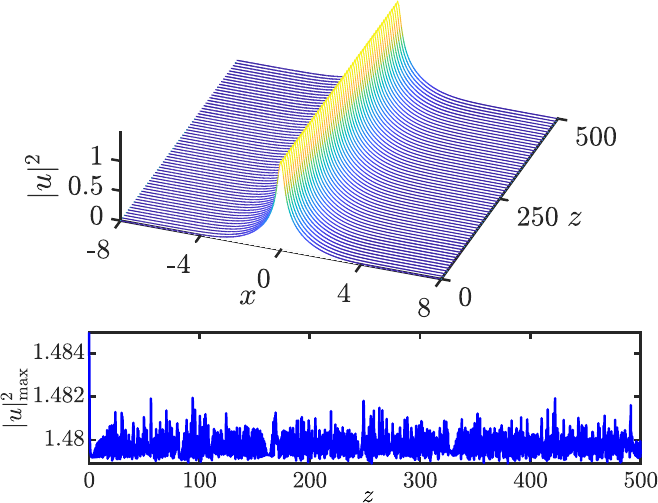}
\caption{Panel (a) and (b) display the evolution of the CW state, as
produced by the simulations of Eq. (\protect\ref{delta}) with $W_{0}=1.5$, $%
g=3$. The initial conditions include the stationary profile, given by Eqs. (%
\protect\ref{sol2}) and (\protect\ref{xi2}) with $k=0.5$, and perturbation (%
\protect\ref{eps}) with $\protect\varepsilon =0.001$ and $p=0.3$.}
\label{Fig16}
\end{figure}

\begin{figure}[h]
\centering
\includegraphics[width=4in]{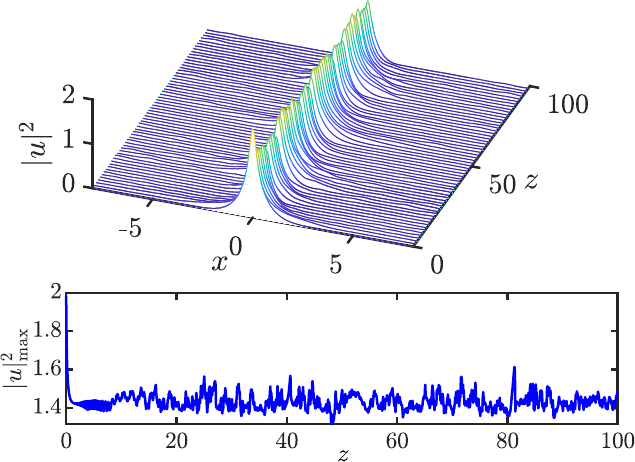}
\caption{The same as in Fig. \protect\ref{Fig16}, but with a much larger
perturbation amplitude, $\protect\varepsilon =0.1$ in Eq. (\protect\ref{eps}%
).}
\label{Fig17}
\end{figure}

An illustration of the stability of the CW with the transverse profile
provided by Eqs. (\ref{sol2}) and (\ref{xi2}), with $W_{0}=1.5$, $g=3$, $%
k=0.5$, is provided by Fig. \ref{Fig16}, in which the simulated propagation
of the CW state, including initial perturbation (\ref{eps}) with $%
\varepsilon =0.001$ and $p=0.3$, is displayed. To additionally test the CW
robustness, a similar simulation is displayed in Fig. \ref{Fig17} for the
perturbation with a much larger amplitude, $\varepsilon =0.1$. It is clearly
observed that the CW is not destroyed even by strong perturbations.

\section{Conclusion and Discussion}

The objective of this work is to consider the dynamics of Q1D
(quasi-one-dimensional) CW (continuous-wave) states and related patterns
(such as chains of 2D solitons) in the optical medium with the CQ
(cubic-quintic) nonlinearity and the trapping potential trough. In the model
with the usual smooth trapping trough, we have reproduced the known MI
(modulational instability) of CW states with the transverse structure
corresponding to the GS (ground state) of the trapping potential, and
produced new results for the CW states with the transverse DM (dipole-mode)
structure, which corresponds to the first excited state of the potential
trap. In particular, it is found that the CWs of both types, GS and DM, are
``practically stable" near edges of their existence domains, which was not
reported in earlier studies. Then, stable periodic chains of 2D solitons
with alternating signs, maintained by the potential trough, have been found
as the stationary manifestation of the MI, and their dynamics, initiated by
the application of localized kicks, was explored.
Another essentially novel result is represented
by two families of exact CW solutions in the model with the delta-functional
trapping potential. A remarkable result is the complete stability of one
family, which does not exist without the trapping potential, because it is
composed of solutions which are singular in the free space.

As an extension of the analysis, it may be interesting to consider chains of
2D solitons with embedded vorticities, identical or alternating (in the
latter case, these are vortex-antivortex chains), trapped in the Q1D
potential. It is known that individual vortex solitons with unitary and
multiple topological charges have their stability regions in the free-space
NLS equation with the CQ nonlinearity \cite{Quiroga,Pego}, therefore the
trapped chain may be stable too. As a new dynamical problem, one may address
collisions between counterpropagating 2D solitons moving in the potential
trough.

It is relevant to compare the results reported in this paper with works
which addressed the configuration with an annular potential trough, instead
of the rectilinear one considered here. In that case, the longitudinal
wavenumber $q$ represents the vorticity and cannot be excluded, cf.
Eqs. (\ref{CW}) and (\ref{k}). As demonstrated theoretically in Ref. \cite{Dong1}, the interplay of
the annular potential trough with the CQ nonlinearity supports stable
vortex states with the winding number (topological charge) $m$ -- at least,
up to $m=12$. As concerns the MI, in this case of the annular potential
trap it takes the form of the azimuthal instability of axially uniform
states, which rearranges them into stable modes periodically structured
along the azimuthal coordinate. The analysis has produced such stable
solutions in the form of multipoles, up to the order $n=20$ (at least) \cite{Dong2}.
The stable chains of 2D solitons with alternating signs, reported
above (see Fig. \ref{Fig10}), are similar to the multipoles trapped in the
annular potential trough.

In the absence of the annular potential trough,
the MI-driven transformation of a ring-shaped beam with
vorticity $q=1$ into a circular cluster of 2D solitons with alternating signs was
reported experimentally in Ref. \cite{Cid2}. In the same work, the experimental results were
reproduced by numerical simulations of the underlying NLS equation. Referring to the same
optical material featuring the CQ nonlinearity, CS$_{2}$, which was used in
work {Cid2}, and the same wavelength and power density of the laser
pump, \textit{viz}., $800$ nm and $\simeq 200$ GW/cm$^{2}$, it is
expected that the setup investigated in the present paper can be implemented
with the rectilinear potential trough of the spatial width $\simeq$ 50 \textmu m,
propagation distance $0<z\lesssim 2$ cm being sufficient for the experiment.

It is relevant to compare the system considered here with other models which
feature competing nonlinearities, similar to the CQ terms in Eq. (\ref{NLS}).
A physically relevant possibility is the nonlinearity which maintains
\textit{quantum droplets} in the two-component BEC \cite{Petrov1,Petrov2,Petrov3}.
In the limit of the 2D system, which is tightly
confined in the third direction, with the corresponding transverse size
smaller than the underlying BEC healing length, the nonlinear term in the
effective 2D Gross-Pitaevskii equation (GPE), which includes the
Lee-Huang-Yan (LHY) \cite{LHY} correction produced by quantum fluctuations
around the mean-field (MF) state, is $|u|^{2}\ln \left(
|u|^{2}/u_{0}^{2}\right) \cdot u$, where $u_{0}^{2}$ is the density at which
the nonlinearity changes its sign from self-attraction to repulsion \cite{Petrov2}.
On the other hand, the LHY-corrected GPE for the loosely confined
2D system, whose transverse size exceeds the heeling length, takes the
following form, in the normalized form \cite{Petrov1}:

\begin{equation}
iu_{t}=-\frac{1}{2}\left( u_{xx}+u_{yy}\right) -\gamma
|u|^{2}u+|u|^{3}u-W(x)u,  \label{GPE}
\end{equation}
where $t$ is the scaled time, $\gamma \geq 0$ is the effective strength of
the MF self-attraction, the quartic term, $|u|^{3}u$, represents the LHY
repulsion, and $W(x)$ is the Q1D trapping potential, which can be
induced by an appropriate laser illumination of the BEC, cf. Eq. (\ref{NLS}).
In particular, in the limit case of the ``LHY superfluid"
\cite{LHY fluid}, $\gamma =0$, with the full cancellation of the MF
interaction, when the GPE nonlinearity is represented solely by the LHY
term, and the potential trap is taken in the same delta-functional form as
in Eq. (\ref{delta}), $W(x)=-\sqrt{\pi }W_{0}\delta (x)$, the exact solution
of Eq. (\ref{GPE}) for the trapped CW state is

\begin{equation}
u\left( x,t\right) =e^{-i\mu t}\frac{\left( -5\mu /2\right) ^{1/3}}{\left[
\sinh \left( 3\sqrt{-\mu /2}\left( |x|+\xi _{\mathrm{LHY}}\right) \right) \right] ^{2/3}},  \label{sol3}
\end{equation}cf. Eq. (\ref{sol2}), with offset\begin{equation}
\xi _{\mathrm{LHY}}=\frac{1}{3\sqrt{-2\mu }}\ln \left( \frac{\sqrt{\pi /8}W_{0}+\sqrt{-\mu }}{\sqrt{\pi /8}W_{0}-\sqrt{-\mu }}\right) ,  \label{xiLHY}
\end{equation}
cf. Eq. (\ref{xi2}). The CW state given by Eqs. (\ref{sol3}) and (\ref{xiLHY})
exists in the interval of values of the chemical potential $-$ $\sqrt{\pi
/8}W_{0}<\mu <0$, and it is obviously immune to the MI, as Eq. (\ref{GPE})
with $\gamma =0$ includes no MI-inducing factor.

\vspace{6pt}
\section*{Acknowledgment}

The work of T.M. is supported by the Faculty of Engineering, Naresuan
University, Thailand (Grant No. R2569E056).

\end{document}